\documentclass{an}
\usepackage{graphicx}
\usepackage{times}
\usepackage{amsmath,amssymb}
\usepackage{natbib}
\bibpunct{(}{)}{;}{a}{}{,}
\newcommand{\ros}{ROSAT}
\newcommand{\chan}{Chandra}
\newcommand{\xmm}{XMM-Newton}
\newcommand{\eROS}{eROSITA}

\newcommand{\nh}{N_{\rm H}}

\def \msev{M7}

\def \jtenfull{\object{2XMM~J104608.7-594306}}

\begin{document}
\Pagespan{789}{}
\Yearpublication{}%
\Yearsubmission{}%
\Month{}%
\Volume{}%
\Issue{}%
\DOI{}%
\title{Follow-up of isolated neutron star candidates from the \eROS\ survey}
\author{A.~M.~Pires\inst{1}
        \fnmsep\thanks{Corresponding author:
        \email{apires@aip.de}\newline}
\and A.~D.~Schwope\inst{1}
\and C.~Motch\inst{2}}
\titlerunning{Follow-up of isolated neutron stars from \eROS}
\authorrunning{Pires, Schwope, \& Motch}
\institute{
     Leibniz-Institut f\"ur Astrophysik Potsdam (AIP), An der Sternwarte 16, 
     14482 Potsdam, Germany
\and Observatoire Astronomique de Strasbourg, Universit\'e de Strasbourg, CNRS, 
     UMR 7550, 11 rue de l'Universit\'e, F-67000 Strasbourg, France}
\received{}
\accepted{}
\publonline{}
\abstract{%
Peculiar groups of X-ray emitting isolated neutron stars, which include magnetars, the ``Magnificent Seven'', and central compact objects in supernova remnants, escape detection in standard pulsar surveys. Yet, they constitute a key element in understanding the neutron star evolution and phenomenology. Their use in population studies in the galactic scale has been hindered by the scarcity of their detection. The all-sky survey of \eROS\ on-board the forthcoming Spectrum-RG mission has the unique potential to unveil the X-ray faint part of the population and constrain evolutionary models. To create a forecast for the four-year all-sky survey, we perform Monte Carlo simulations of a population synthesis model, where we follow the evolutionary tracks of thermally emitting neutron stars in the Milky Way and test their detectability. In this work, we discuss strategies for pinpointing the most promising candidates for follow-up observing campaigns using current and future facilities.%
}
\maketitle
\section{Introduction}
\begin{figure}[t]
\begin{center}
\includegraphics*[width=0.485\textwidth]{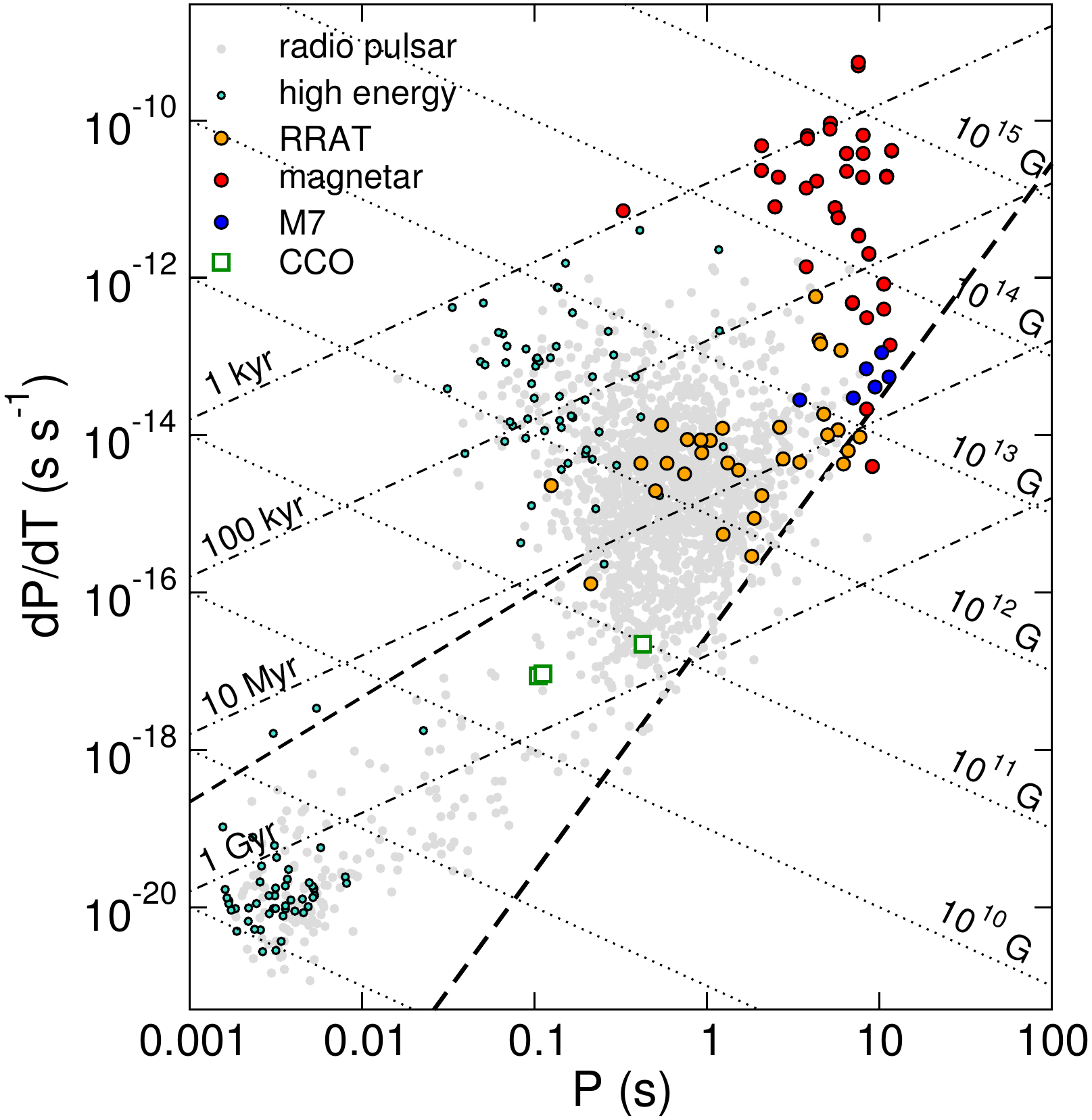}
\end{center}
\caption{Pulsar spin period as a function of spin down ($P$\,--$\dot{P}$) diagram of the galactic population of neutron stars. 
\label{fig_pdotp}}
\end{figure}

Since the discovery of the first radio pulsar nearly fifty years ago \citep{hew68}, the population of known neutron stars in our Galaxy has grown to over 2500 \citep[][see the $P$\,--\,$\dot{P}$ diagram of Figure~\ref{fig_pdotp}]{man05}. Classical (rotation-powered) and millisecond (recycled) pulsars, as well as a growing number of bursting radio sources, dubbed rotating radio transients \citep[RRATs,][]{lau06}, make up most of the population. 
Only a small fraction of the observed sample consists of peculiar groups of X-ray emitting isolated neutron stars (INSs), which escape detection in standard radio and $\gamma$-ray pulsar surveys \citep[for an overview, see][]{mer11a,har13}.
These are the young and energetic magnetars \citep[e.g.][for a review]{mer15}; the group of nearby, middle-aged, thermally emitting sources known as the Magnificent Seven \citep[\msev,][]{tur09}; and the young and seemingly weakly magnetised central compact objects in supernova remnants \citep[CCOs, also dubbed `anti-magnetars';][]{got13a}. 

Despite their scarcity, the peculiar groups are key to understanding evolutionary aspects that are not predicted by theory, nor probed by the normal pulsar population. In particular, scenarios involving magnetic field decay \citep[e.g.][]{agu08,pon09,vig13} and field burial by fallback accretion after the supernova explosion \citep[e.g.][]{che89,gep99,how11} are worth noting, as they are expected to signficantly alter the star's rotational and thermal evolution and, hence, visibility across the electromagnetic spectrum. Therefore, to survey our Galaxy for X-ray emitting neutron stars in the hope of discovering evolutionary missing links is likewise important for a thorough understanding of their physics and phenomenology.
 
The Extended R\"ontgen Survey with an Imaging Telescope Array \citep[\eROS;][]{pre17} is the primary instrument on the new Russian Spectrum-RG mission, which is planned for launch in September 2017. In the first four years of the mission, \eROS\ will survey the X-ray sky with unprecedented sensitivity, spectral and angular resolution, which is a timely opportunity for a better sampling of neutron stars that are especially silent in the radio and $\gamma$-ray regimes. 
The selection of newly proposed INS candidates among the myriad of \eROS-detected sources and transient phenomena will be challenging: their identification and characterisation will require synergy with multiwavelength, large-scale, photometric and spectroscopic surveys that are operational in the near future \citep{mer12}. Moreover, state-of-the-art observing facilities, in particular 8-m-class optical telescopes, the \xmm\ and \chan\ Observatories, are crucial for dedicated follow-up campaigns to be conducted already in the immediate aftermath of the \eROS\ survey. 

To estimate the number of INSs to be detected in the forthcoming \eROS\ All-Sky Survey (eRASS) through their thermal X-ray emission, we performed Monte Carlo simulations of a population synthesis model, which is described in Section~\ref{sec_popsyn}. The results are presented in Section~\ref{sec_results}. In Section~\ref{sec_followup} we outline strategies for selecting the most promising candidates for follow-up observing campaings, focusing on the time frame of the first years after the survey. The summary and perspectives are presented in Section~\ref{sec_conclusions}.
\section{Population synthesis\label{sec_popsyn}}
Population synthesis of thermally emitting INSs \citep[see e.g.][for a review]{pop07a} relies on the parametrisation of unknown properties inherited at birth, which are relevant to the cooling development and source emissivity. Together with realistic descriptions of the galactic gravitational potential, the interstellar medium (ISM) distribution, and the detector and survey characteristics, these ingredients determine the neutron star evolution and detectability. 

To create a forecast for \eROS, we developed a model in which neutron stars are created from a progenitor population of massive stars distributed in the spiral arms of the galactic disk; after the supernova explosion and the imparted `kick' velocity, their evolution in the galactic gravitational potential is followed while they cool down emitting soft X-rays. The expected source count rates and total flux are then computed in the \eROS\ detectors, taking into account the absorbing material in the line-of-sight and the celestial exposure after four years of all-sky survey. For a detailed description of the main ingredients of the model and simulation procedure, see \citet{pir09c}; Pires, Schwope \& Motch (in preparation).
\subsection{Galactic model and neutron star birth properties\label{sec_nsbirth}}
As a starting point, we adopt a galactic model similar to that described in \citet{fau06}. We reproduce the spiral arm structure of the Milky Way by invoking four parametrised logarithmic arms \citep{wai92}. For the gravitational potential, we consider the contributions of the disk, bulge, and halo, and also take into account spiral wave perturbations on the exponential disk \citep[see][for references]{pir09c}. For the ISM, we adopt an analytical distribution of the absorbing material based on layers of hydrogen in atomic and molecular form \citep{dic90,boe91}. The cross section in \citet{mor83} is assumed for the photoelectric absorption. 

The neutron stars in our simulations are created one by one as a function of galactocentric radius. For each source, the azimuthal angle is determined so that the neutron star's original position, projected on the plane, intersects one of the four spiral arms. 
A vertical distance relative to the plane is assigned so that the ensemble of sources is exponentially distributed above and below the disk, with a scale height of 50\,pc.
The radial distribution of neutron star progenitors follows the one suggested by \citet{yus04} and is normalised for a birthrate of 2.1\,century$^{-1}$ \citep[assumed to be constant during the timescale relevant for cooling, i.e.~$\ll10^8$\,yr;][]{gil07}.
\subsection{Neutron star evolution\label{sec_nsevol}}
After the supernova explosion, the newly born neutron star receives a kick with no preferential direction and a three-dimensional speed that is exponentially distributed with a mean of 380\,km\,s$^{-1}$ \citep{fau06}. The star dynamically evolves in the galactic potential for a time equivalent to its age. The hydrogen column density towards the source is then quantified by integrating the amount of intervening material in the line-of-sight. 

The source count rate $S$ in a given energy band $\epsilon_1-\epsilon_2$ is computed for each neutron star as a function of its present-day distance $d$ and temperature $T^\infty$ (assuming a cooling curve; see below), the calculated column density $\nh$ towards the source, the photon cross section $\sigma_\epsilon$, and the combined effective area $A_\epsilon$ of the \eROS\ detectors:

\begin{equation}
S = \int_{\epsilon_1}^{\epsilon_2} \frac{f_\epsilon}{\epsilon}A_\epsilon\exp\Big[-\sigma_\epsilon \nh(r=d,l,b)\Big]\,d\epsilon
\label{eq_cr}
\end{equation}

\noindent where $f_\epsilon=(R_\infty/d)^2F_\epsilon^\infty$ is the geometrically diluted flux, $R_\infty$ is the redshifted emission radius, and $F_\epsilon^\infty$ is the source's blackbody flux at infinity, assumed isotropical.

\begin{figure}[t]
\begin{center}
\includegraphics*[width=0.485\textwidth]{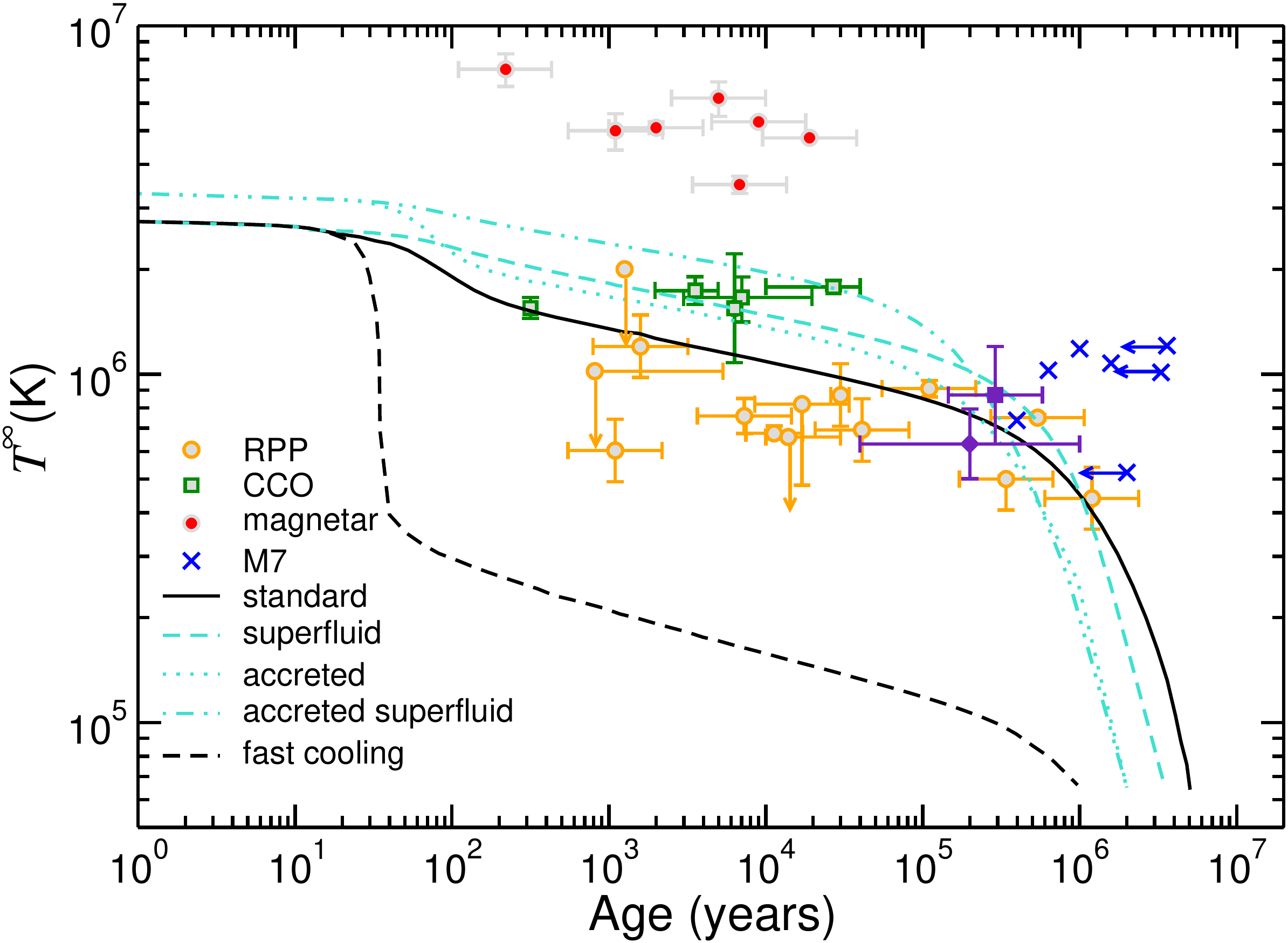}
\end{center}
\caption{Cooling-age diagram for different INS groups (data points; see legend. Details can be found in \citealt{pir15}). 
The standard neutrino candle is shown as a solid black line; effects of proton superfluidity in the core and of a light-element accreted envelope are shown in turquoise. The enhanced-cooling scenario (associated with sufficiently massive neutron stars) is illustrated by the dashed black line.\label{fig_ccurve}}
\end{figure}

For simplicity, we assume that all synthetic stars have a (non-exotic) nucleon core, canonical mass and radius (1.4\,M$_\odot$ and 12\,km), and cool down slowly as the so-called standard neutrino candle \citep[][shown as a solid black line in Figure~\ref{fig_ccurve}]{yak11}. In fact, the star's mass, magnetic field and its decay, the presence of an accreted envelope, among others, play a crucial role at determining the present-day luminosity of the neutron star. This is evidenced by the data points in the cooling-age diagram of Figure~\ref{fig_ccurve} that show inconsistent temperatures with respect to those of the standard scenario. These aspects of cooling, as well as their impact on the number of INSs to be detected by the eRASS, will be investigated elsewhere (Pires, Schwope, \& Motch, in preparation).
\subsection{Survey properties\label{sec_survey}}
The \eROS\ mirror system consists of seven identical modules \citep[for details, see][]{pre14b}. The observatory is planned to scan the X-ray sky eight times in four years; in each scan, a source moves on a track on the detector plane. Due to the vignetting of the point-spread function away from the aimpoint, the combined effective area of the seven detectors has to be averaged over the entire field-of-view of the telescope ($1.03^\circ$ diameter), in order to compute count rates during the survey phase.

The exact instrument response and energy passband will be determined by the properties of the optical blocking filters and on the available telemetry. Five of the seven detectors have a 200\,nm aluminium on-chip filter which, depending on the environmental conditions, might be operated with or without an additional external 200\,nm polyimide filter (K.~Dennerl, P.~Predehl, private communication\footnote{The \eROS\ calibration files are available at https://wiki.mpe.mpg.de/eRosita/erocalib\_calibration.}). The two remaining \eROS\ detectors have no on-chip filter, thus requiring a 200\,nm polyimide filter, coated with 100\,nm aluminium, to be placed at the filter wheel to block the contaminating light at any circumstances.

\begin{figure}[t]
\begin{center}
\includegraphics*[width=0.485\textwidth]{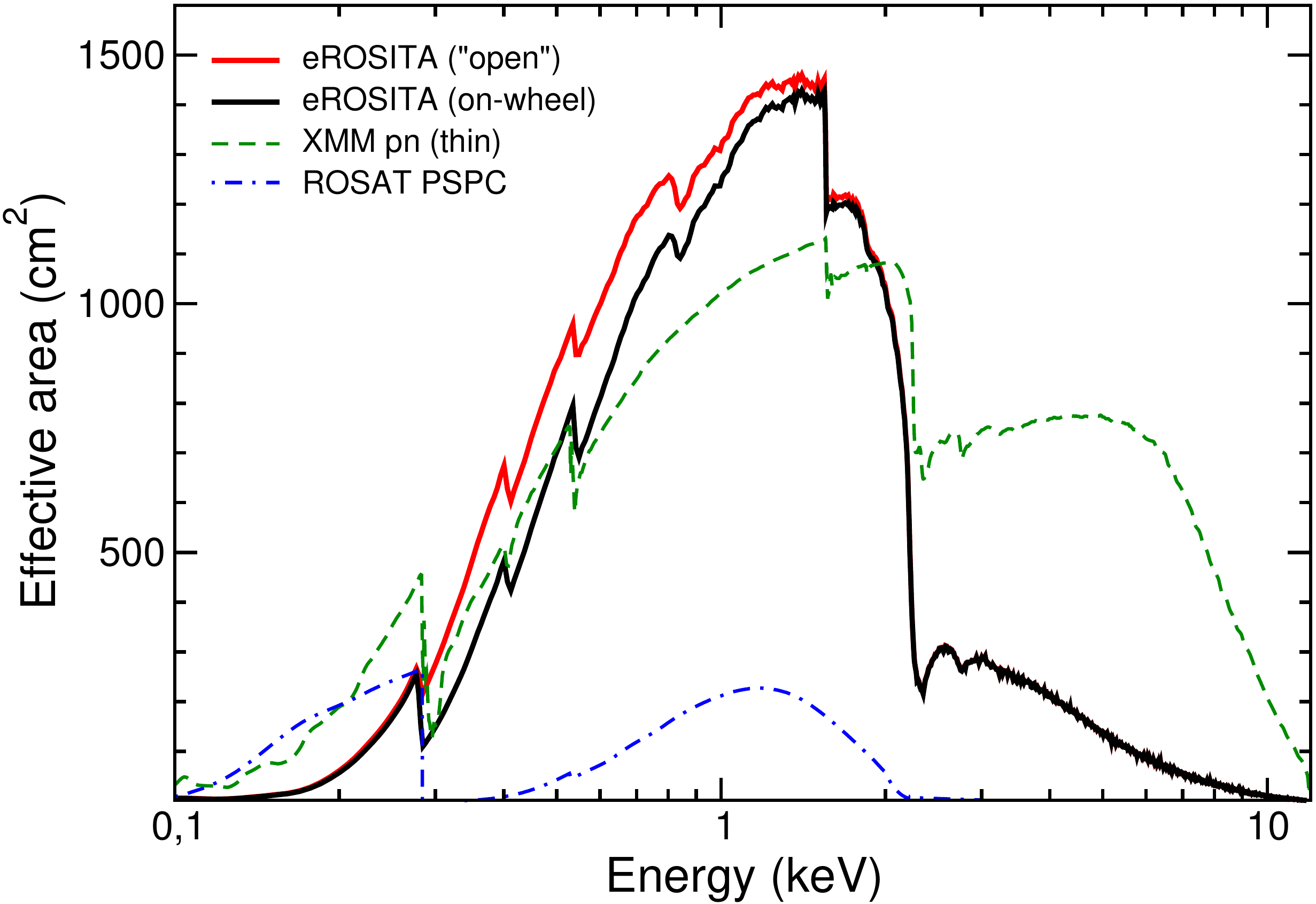}
\end{center}
\caption{Total effective area of \eROS\ (seven mirror modules) averaged over the field-of-view for two possible filter configurations, in comparison with those of other X-ray missions (see legend and text). 
\label{fig_effarea}}
\end{figure}

Accordingly, we considered in the simulations two possible configurations of optical blocking filters: the `open' option, where no additional filter is employed for the five detectors with on-chip filter, and the most conservative `on-wheel' option, where all detectors have external filters placed on the wheel. 
The resulting total effective area of the seven telescopes, averaged over the field-of-view, can be seen in Figure~\ref{fig_effarea}. 
For comparison, we also show in Figure~\ref{fig_effarea} the effective areas of the \xmm\ EPIC-pn and \ros\ PSPC detectors.

Assuming an observing efficiency of 100\%, an average exposure of $t_{\rm flat}=2.5$\,ks will be reached after four years \citep{mer12}. This number is independent of the exact scanning law and can be considered as a flat all-sky exposure. 
In reality, the celestial exposure will be determined by the survey strategy. As the scanning axis of the satellite will be oriented approximately towards the Sun, the exposure depths will be unevenly distributed, where the lowest (highest) exposures will be close to the ecliptic plane (at the poles). 
We thus adopted in the simulations an exposure map that mimicks this behaviour (Figure~\ref{fig_expmap}). The exact scanning law, which is to-date being discussed within the \eROS\ consortium, will further introduce a longitudinal pattern due to the non-uniform movement of the spacecraft within the ecliptic; this will consequently redistribute the exposure at higher latitudes and around the poles (J.~Robrade, private communication). 
\section{Thermal INSs in the eRASS\label{sec_results}}
\begin{figure}[t]
\begin{center}
\includegraphics*[width=0.485\textwidth]{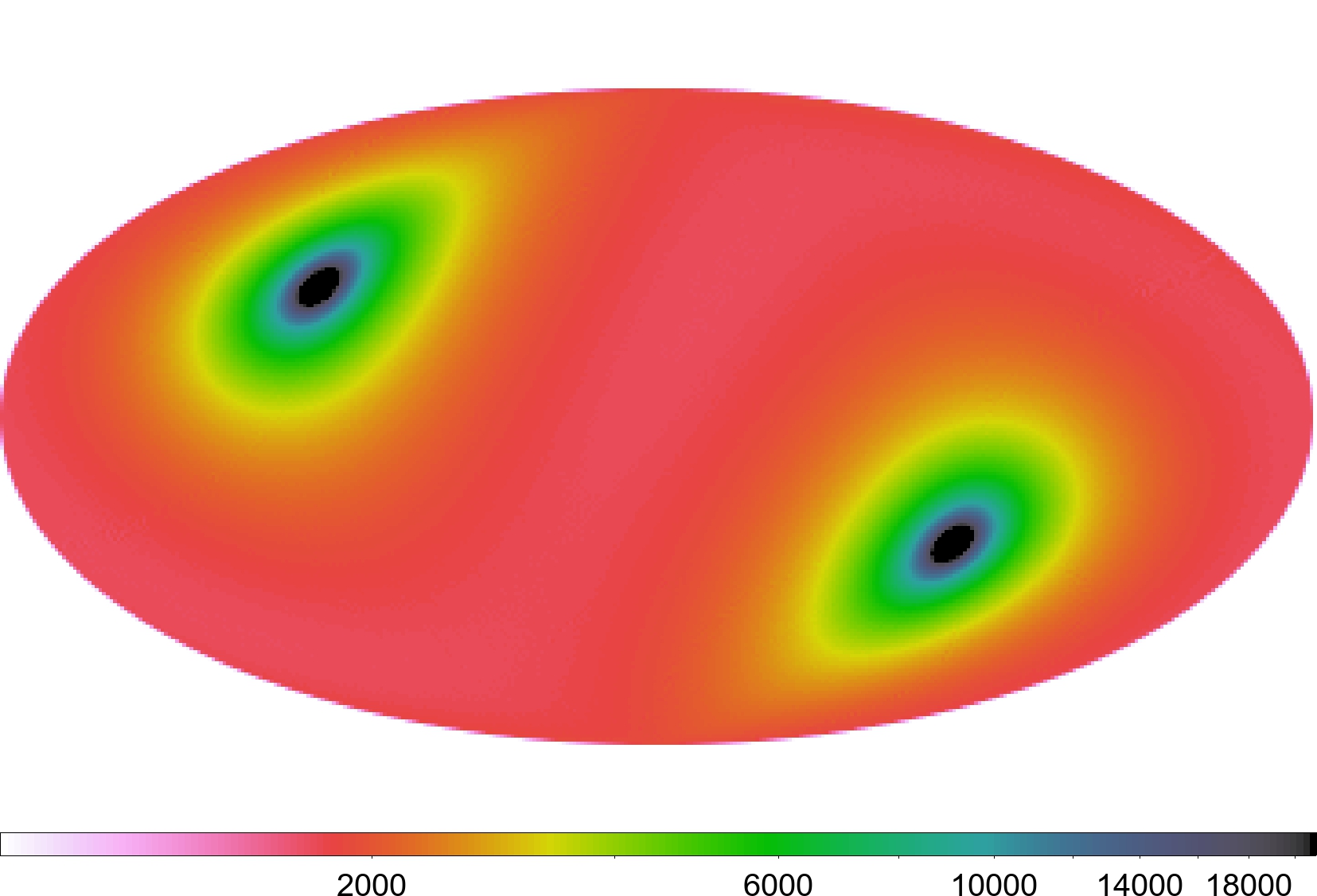}
\end{center}
\caption{Exposure map of the eRASS adopted in the population synthesis, in galactic coordinates. The colour-coded bar shows the exposure in seconds, which accumulates around the ecliptic poles (credit: J.~Robrade).
\label{fig_expmap}}
\end{figure}

Independent runs of Monte Carlo simulations of the population synthesis model are performed for each filter configuration. 
In each run, the integral in equation (\ref{eq_cr}) is solved for the synthetic neutron stars in five pre-defined energy bands, which are adopted for convenience as those of the \xmm\ source catalogue\footnote{That is, 0.2\,--\,0.5\,keV, 0.5\,--\,1\,keV, 1\,--\,2\,keV, 2\,--\,4.5\,keV, and 4.5\,--\,12\,keV; in practice, count rates from thermal INSs are insignificant above 2\,keV. The low end of the energy band will be determined by telemetry constraints (P.~Predehl, private communication).}. 
Total detected counts are then computed taking into account the accumulated exposure towards each star (Figure~\ref{fig_expmap}); the sources with more than 30 detected counts ($0.2-2$\,keV) are collected for each simulation and consist of our working `observed' sample. Technically, the adopted count limit ensures a low probability of the observed flux being caused by a random background fluctuation ($p\sim4.5\times10^{-5}$; G.~Lamer, private communication).

Over 50 Monte Carlo realisations of each filter configuration, we obtain on average $85\pm3$ (on-wheel) to $95\pm3$ (open) thermally emitting INSs to be detected by the eRASS after four years (errors are $2\sigma$ confidence levels). The 11\% reduction on sample size of the more conservative filter option is due to the loss of effective area at low energies (up to 1.5\,keV; Figure~\ref{fig_effarea}). Similarly, significant losses in the number of observed sources are to be expected if the low end of the energy band is considerably constrained above 0.2\,keV for telemetry reasons (of 3\%, 11\%, and 28\% for 0.3\,keV, 0.4\,keV, and 0.5\,keV, respectively; in every case, the same lower limit of 30 deteted counts was considered). 

In Table~\ref{tab_aver} we list the average or median properties\footnote{We list the number of detected sources $N$, limiting count rate $S_{\rm min}$, observed flux $f_{\rm X}$, percentage of neutron stars brighter than $f_{13}$, blackbody temperature $kT$, distance $d$, column density $\nh$, age, three-dimensional velocity $v$, and average exposure $t_{\rm exp}$ (see text for details).} of the observed INS samples, for each filter configuration. On average, the more conservative filter option detects slightly brighter, hotter, slower, and closer-by neutron stars. The differences are however only significant within the statistical errors of the Monte Carlo simulations and will not reflect real measurement precisions. 

\begin{figure}[t]
\begin{center}
\includegraphics*[width=0.475\textwidth]{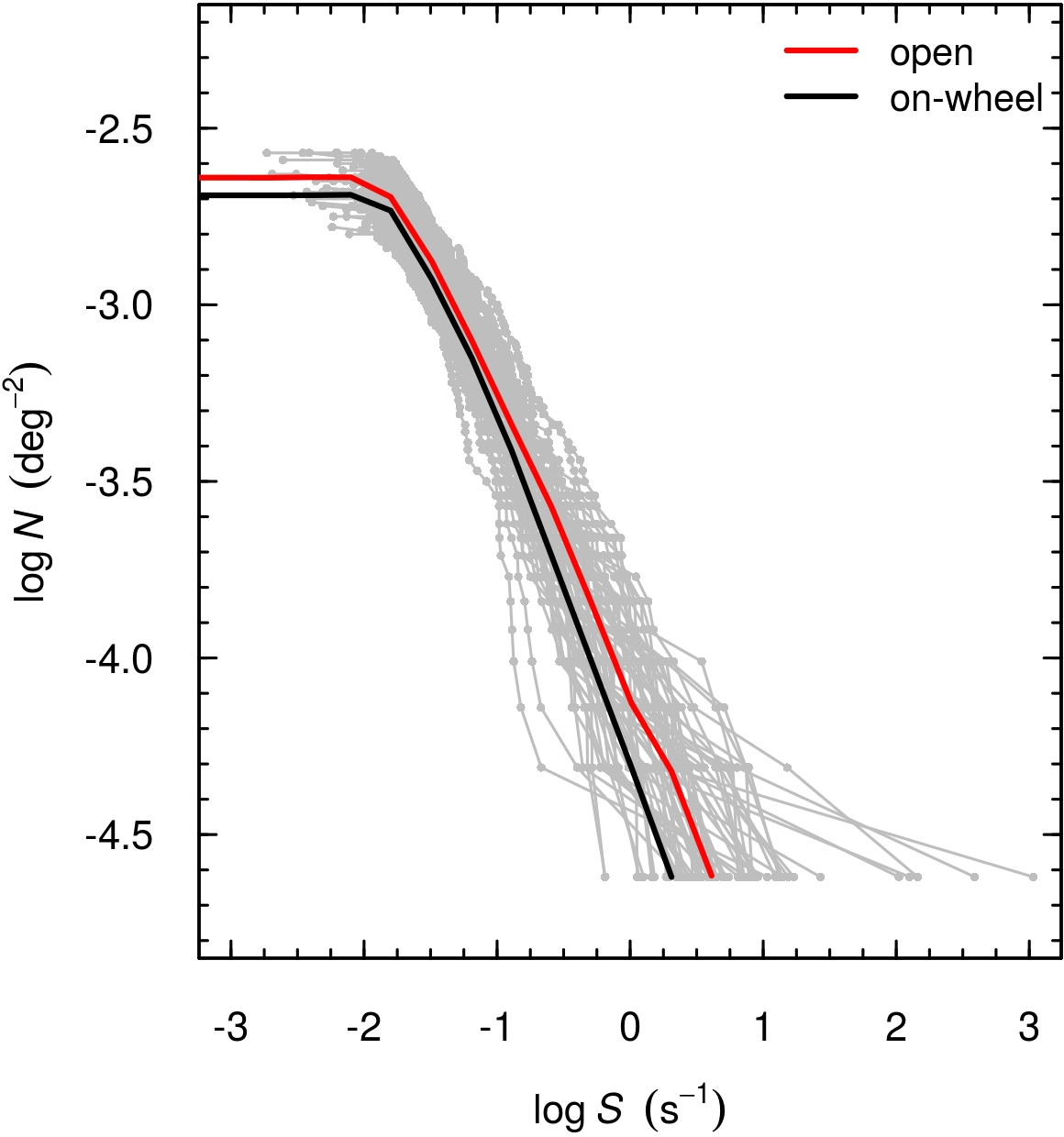}
\end{center}
\caption{All-sky $\log N$\,--$\log S$ curves (0.2-2\,keV). Red and black solid curves are the average of 50 curves for the open and on-wheel filter configurations, respectively (grey lines are the individual results of the simulations).
\label{fig_lnls}}
\end{figure}

Consistently between the two filter options, distances are within 400\,pc and 8\,kpc; around 60\% of the sources are within 5\,degrees of the galactic disk, whereas 20\% of the sources undergo column densities higher than $\nh=10^{22}$\,cm$^{-2}$. The accumulated survey exposure towards the sources ranges from 1.6\,ks to 8\,ks, with a median around 1.9\,ks; 23\% of the sources receive a higher exposure than $t_{\rm flat}$ (Sect.~\ref{sec_survey}). Finally, the observed flux of the faintest neutron stars is around $10^{-14}$\,erg\,s$^{-1}$\,cm$^{-2}$ and the median is five times this value; 28\% of the sources are found to be brighter than $f_{13}=10^{-13}$\,erg\,s$^{-1}$\,cm$^{-2}$. Considering the flux limit of the \ros\ Bright Source Catalogue ($5\times10^{-13}$\,erg\,s$^{-1}$\,cm$^{-2}$; \citealt{vog99b}), the all-sky survey of \eROS\ will be approximately 50 times more sensitive to neutron stars than that of its predecessor. In Figure~\ref{fig_lnls} we show the resulting number of detected sources per square degree on the sky as a function of \eROS\ count rates ($\log N$\,--\,$\log S$ curves).
\begin{table}[t]
\begin{center}
\caption{Average/median properties of the observed INS sample\label{tab_aver}}
\begin{tabular}{l l l l}
\hline\hline
 & open & on-wheel & \\
\hline
$N$ & $95\pm3$ & $85\pm3$ & \\
$S_{\rm min}$ & $8.2(4)\times10^{-3}$ & $8.9(4)\times10^{-3}$ & s$^{-1}$ \\
$f_{\rm X}^{\rm min}$ & $0.93(7)\times10^{-14}$ & $1.27(7)\times10^{-14}$ & [cgs]\\
$f_{\rm X}^{\rm med}$ & $4.349(26)\times10^{-14}$ & $5.30(6)\times10^{-14}$ & [cgs] \\
$\ge f_{13}$ & 26 & 31 & \% \\
$kT_{\rm min}$ & $43.1\pm1.5$ & $45.0\pm1.6$ & eV \\
$kT_{\rm max}$ & $190\pm10$ & $180\pm7$ & eV \\
$kT_{\rm med}$ & $91.84(14)$ & $92.66(20)$ & eV \\ 
$d_{\rm min}$ & $360\pm40$ & $390\pm40$ & pc \\
$d_{\rm med}$ & $1.789(6)$ & $1.707(6)$ & kpc \\
$N_{\rm H}^{\rm med}$ & $4.608(24)\times10^{21}$ & $4.60(3)\times10^{21}$ & cm$^{-2}$ \\
age & $1.99(6)\times10^5$ & $1.89(8)\times10^5$ & years\\
$v_{\rm med}$ & $413.9\pm2.5$ & $406.3\pm2.9$ & km\,s$^{-1}$ \\
$t_{\rm exp}$ & $2.23\pm0.04$ & $2.22\pm0.03$ & ks \\ 
\hline
\end{tabular}
\end{center}
\end{table}
\section{Strategies for follow-up\label{sec_followup}}
The selection of INS candidates from the eRASS will primarily rely on cuts in X-ray colour space and cross-correlations with catalogues at other wavelengths. The main selection criteria, which exclude approximately 95\% of the X-ray sources in a flux-limited sample \citep[see e.g.][for details]{pir09b}, require a soft energy distribution (for instance, in consistency with the positions of absorbed blackbody templates in hardness ratio diagrams) and the absence of catalogued optical and infrared counterparts, usually down to visual magnitudes 21 to 23. 

The identification of thermally emitting INSs at faint fluxes is challenged further by the ISM extinction and source confusion in the crowded sky regions
of the galactic plane. Based on positional coincidence, cross-correlation procedures tend to assign unrelated optical or near-infrared sources as the counterpart of a given X-ray source, thus erroneously excluding it as a potential INS candidate \citep[see e.g.][]{rut03}. 
In this context, the use of statistical multiwavelength cross-identification methods is key to optimise source characterisation and minimise the number of false positives \citep[see e.g.][]{pin16}. 

Visual screening and dedicated follow-up observations are then necessary to filter out remaining spurious detections (for instance, due to out-of-time events), as well as other classes of soft X-ray emitters that happened to satisfy our selection criteria, such as faint polar-type cataclysmic variable systems (CVs) and active galactic nuclei (AGN). In this respect, the intrinsic INS faintness in the visible waveband, with typical X-ray-to-optical flux ratios in excess of $10^4$, comes in handy: at the flux level probed by the eRASS, the optical counterparts of newly proposed candidates will be virtually impossible to detect with current facilities ($m_V>27$). On the other hand, CVs and AGN have much less extreme flux ratios \citep[of less than 300; e.g.][]{schwope99} and are hence easily revealed with 4\,m and 8\,m-class telescopes. Therefore, the absence of counterparts in the X-ray error circle will be a strong evidence that the investigated source is indeed a neutron star.
\begin{figure}[t]
\begin{center}
\includegraphics*[width=0.485\textwidth]{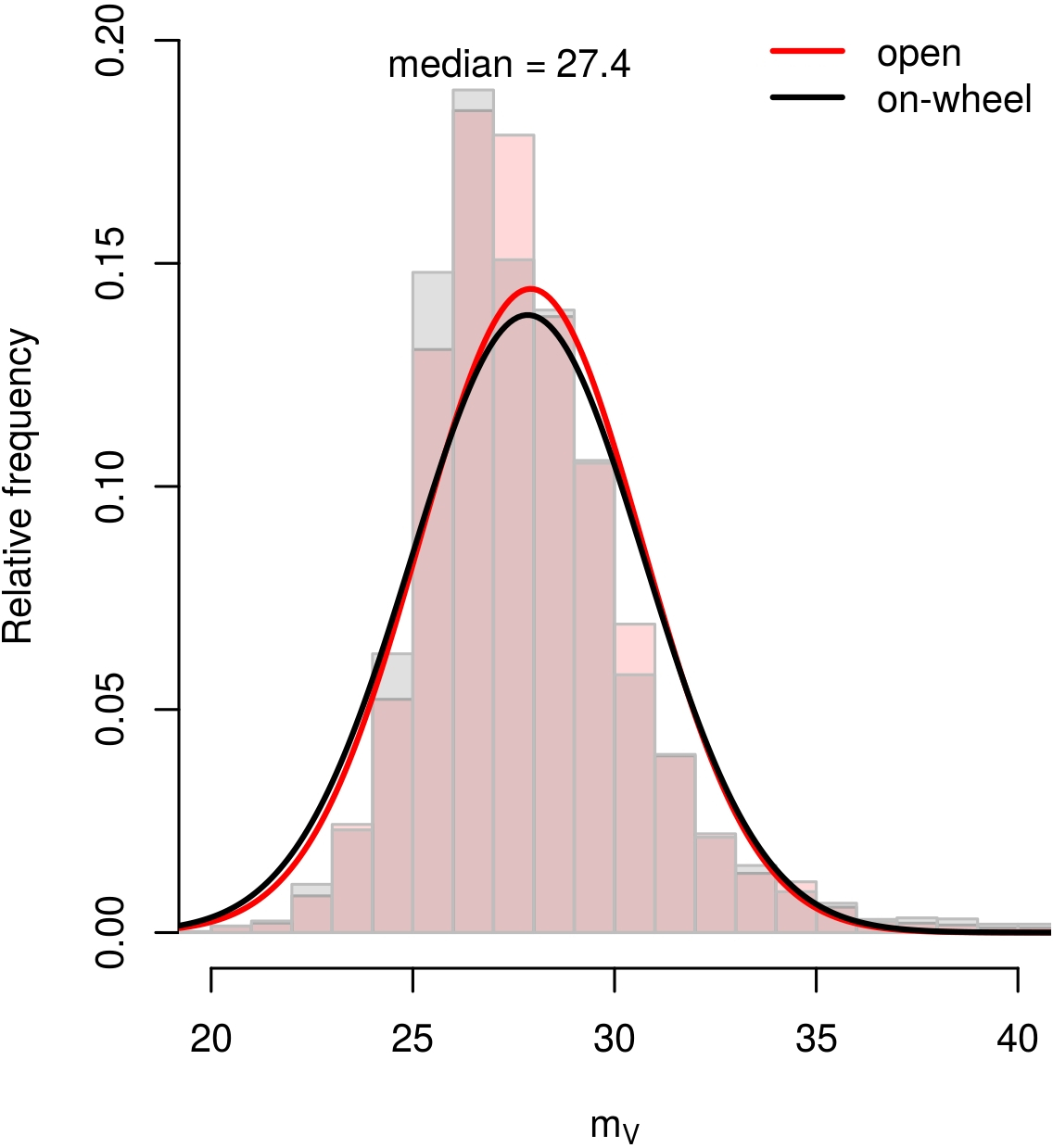}
\end{center}
\caption{Histogram of the optical V magnitudes required for ruling out AGN and CVs among INS candidates. A minimum X-ray-to-optical flux ratio of $F_{\rm X}/F_{\rm V}>10^{3.5}$ is assumed for the \eROS-detected sample of neutron stars. 
\label{fig_magV}}
\end{figure}

Using our simulations, we can derive the optical limit required for ruling out other classes of X-ray emitters for each \eROS\ INS. Assuming a conservative X-ray-to-optical flux ratio of at least $10^{3.5}$ and taking into account the absorption by the ISM, the median $V$ magnitude is $m_{\rm V}=27.44(6)$, with no significant differences between the two filter options (Figure~\ref{fig_magV}). Therefore, excluding the bright end of the population which was already probed by \ros, around 27\% of the \eROS\ INSs are still sufficiently bright in X-rays to be identified and investigated in the years following the \eROS\ survey. 
Interestingly, these sources will be at a similar flux level (on average, $f_{\rm X}=1.26(29)\times10^{-13}$\,erg\,s$^{-1}$\,cm$^{-2}$) than that of the only thermally emitting INS identified to-date in non-\ros\ data, \jtenfull\ \citep{pir12,pir15}. In this regard, the fact that the \xmm\ and \chan\ observatories show good prospects for the next decade of operation is timely news for neutron star physics as well.

Of central importance for a comprehensive understanding of the INS population is the investigation of the evolutionary state of \eROS-discovered sources. In particular, the role played by the decay of the magnetic field in the cooling evolution and source visibility cannot be overlooked, as it partially explains the observed neutron star diversity. However, even the state-of-the-art models are built over particularly uncertain assumptions \citep{vig12b}; moreover, the observed population, including radio and X-ray pulsars, is not sufficient to constrain alternative models of field decay \citep{gul15}. Likewise, while the phenomenology of CCOs has triggered a lively debate over their formation and fate, no attempt to our knowledge has been made in population synthesis to include anti-magnetars as a possible outcome of neutron star evolution. As long as only the X-ray bright end of the radio-quiet INS population is known, the observational and theoretical advances seen in recent years may be brought to a halt. 

The knowledge of the INS spin period and magnetic field is the first step to classify new sources amidst the several families, and populate the $P-\dot{P}$ diagram of Figure~\ref{fig_pdotp} with \eROS-detected INSs. At the intermediate flux level of $10^{-13}$\,erg\,s$^{-1}$\,cm$^{-2}$ and moderate ISM absorption ($\nh\lesssim5\times10^{21}$\,cm$^{-2}$), around 100\,ks exposure with \xmm\ per target will constrain pulsations down to $10\%$ to $15\%$. Spectral absorption lines, which can then be related to the star's magnetic field on the surface, can also be detected at energies above $0.5$\,keV. While snapshot observations with \chan\ will be desirable for locating the X-ray source with sub-arcsecond precision, which is important for sources at low galactic latitudes, longer exposures will allow searching for diffuse emission around the neutron star (for instance, due to a bow-shock or a pulsar-wind nebula). 

Finally, searches for radio and $\gamma$-ray counterparts (for example, around candidate periodicities from X-rays) are essential to unambiguously classify the source among the neutron star families. Pulsar surveys based on e.g.~Fermi-LAT data and the Square Kilometre Array (SKA) precursor MeerKAT, which will become operational in June 2017, will be used. In the long term, the next generation of multiwavelength observing facilities, the European Extremely Large Telescope, the Large Synoptic Survey Telescope, Athena, and SKA, will allow even the faintest \eROS\ neutron stars to be identified and studied in detail.
\section{Summary and conclusions\label{sec_conclusions}}
We developed a model for the population of neutron stars in our Galaxy to test their detectability by the next X-ray survey mission \eROS.
We take into account the absorption of X-ray photons by the interstellar medium, the survey properties, and the celestial exposure. Our study shows that the \eROS\ survey will be sensitive to neutron stars as faint as $f_{\rm X}\sim10^{-14}$\,erg\,s$^{-1}$\,cm$^{-2}$ (0.2\,--\,2\,keV), and is expected to detect a number of 85 to 95 sources in the sky -- a significant increase with respect to \ros. 

Based on X-ray-to-optical flux ratios, optical follow-up will require deep observations to rule out X-ray emitters other than INSs. In particular, the identification of the faintest neutron star candidates will have to wait for the next generation of extremely large telescopes. Nonetheless, sources at intermediate fluxes ($f_{\rm X}\sim10^{-13}$\,erg\,s$^{-1}$\,cm$^{-2}$) can be selected for follow-up investigations using current facilities, in particular 8\,m-class optical telescopes, and the \xmm\ and \chan\ Observatories. Based on that, we anticipate a number of up to 25 discoveries already in the first years following the survey. 

Beyond the discovery of new sources and long-sought evolutionary missing links, the \eROS\ survey has the unique potential to unveil the faint X-ray end of the neutron star population. A better sampling of the sources that are silent in the radio and $\gamma$-ray regimes is essential to probe the population as a whole, constrain the galactic rate of supernova type-II explosions, test alternative evolutionary scenarios involving field decay and fallback accretion, and help us to have a comprehensive view of the neutron star phenomenology.
\acknowledgements
We acknowledge the \eROS\ German Consortium for providing supporting material on various aspects of the mission. We thank the anonymous referee for suggestions that helped improving the manuscript. The work of A.M.P.~is supported by the Deutsches Zentrum f\"ur Luft- und Raumfahrt (DLR) under grant 50 OR 1511.
\vfill
\begingroup
\let\clearpage\relax
%
%
%
%


\def\ref@jnl{}

\def\aj{\ref@jnl{AJ}}                   
\def\araa{\ref@jnl{ARA\&A}}             
\def\apj{\ref@jnl{ApJ}}                 
\def\apjl{\ref@jnl{ApJ}}                
\def\apjs{\ref@jnl{ApJS}}               
\def\ao{\ref@jnl{Appl.~Opt.}}           
\def\apss{\ref@jnl{Ap\&SS}}             
\def\aap{\ref@jnl{A\&A}}                
\def\aapr{\ref@jnl{A\&A~Rev.}}          
\def\aaps{\ref@jnl{A\&AS}}              
\def\azh{\ref@jnl{AZh}}                 
\def\baas{\ref@jnl{BAAS}}               
\def\jrasc{\ref@jnl{JRASC}}             
\def\memras{\ref@jnl{MmRAS}}            
\def\mnras{\ref@jnl{MNRAS}}             
\def\pra{\ref@jnl{Phys.~Rev.~A}}        
\def\prb{\ref@jnl{Phys.~Rev.~B}}        
\def\prc{\ref@jnl{Phys.~Rev.~C}}        
\def\prd{\ref@jnl{Phys.~Rev.~D}}        
\def\pre{\ref@jnl{Phys.~Rev.~E}}        
\def\prl{\ref@jnl{Phys.~Rev.~Lett.}}    
\def\pasp{\ref@jnl{PASP}}               
\def\pasj{\ref@jnl{PASJ}}               
\def\qjras{\ref@jnl{QJRAS}}             
\def\skytel{\ref@jnl{S\&T}}             
\def\solphys{\ref@jnl{Sol.~Phys.}}      
\def\sovast{\ref@jnl{Soviet~Ast.}}      
\def\ssr{\ref@jnl{Space~Sci.~Rev.}}     
\def\zap{\ref@jnl{ZAp}}                 
\def\nat{\ref@jnl{Nature}}              
\def\iaucirc{\ref@jnl{IAU~Circ.}}       
\def\aplett{\ref@jnl{Astrophys.~Lett.}} 
\def\apspr{\ref@jnl{Astrophys.~Space~Phys.~Res.}}
\def\bain{\ref@jnl{Bull.~Astron.~Inst.~Netherlands}} 
\def\fcp{\ref@jnl{Fund.~Cosmic~Phys.}}  
\def\gca{\ref@jnl{Geochim.~Cosmochim.~Acta}}   
\def\grl{\ref@jnl{Geophys.~Res.~Lett.}} 
\def\jcp{\ref@jnl{J.~Chem.~Phys.}}      
\def\jgr{\ref@jnl{J.~Geophys.~Res.}}    
\def\jqsrt{\ref@jnl{J.~Quant.~Spec.~Radiat.~Transf.}}
\def\memsai{\ref@jnl{Mem.~Soc.~Astron.~Italiana}}
\def\nphysa{\ref@jnl{Nucl.~Phys.~A}}   
\def\physrep{\ref@jnl{Phys.~Rep.}}   
\def\physscr{\ref@jnl{Phys.~Scr}}   
\def\planss{\ref@jnl{Planet.~Space~Sci.}}   
\def\procspie{\ref@jnl{Proc.~SPIE}}   

\let\astap=\aap
\let\apjlett=\apjl
\let\apjsupp=\apjs
\let\applopt=\ao

\bibliographystyle{aa}
\bibliography{pires}
\endgroup
\end{document}